**Impact of Gold Prices on Stock Exchange: An Empirical Case Study of Nepal**

Aneel Bhusal* and Madhu Sudan Gautam**
*Institute for Frontier Studies, Nepal
**Kathmandu University School of Education, Nepal

**Abstract**
The purpose of this study is to examine the long-run relationship between gold prices and Nepal Stock Exchange (NEPSE). The statistical techniques used for this study includes Unit Root Augmented Dickey Fuller test, Phillips-Perron, Johnson Co-integration and Granger's Causality tests to measure the long-run relationship between gold prices, NEPSE using monthly data from 1$^{st}$ September 2010 to 31$^{st}$ June 2015. Both NEPSE index and gold price didn't have stationarity at order zero. Findings of the co-integration test indicated that no long-run relationship exist between monthly average gold prices and NEPSE stock index. Results of Granger causality test demonstrated that no causal relationship exists among average gold prices and NEPSE stock indices.

**Main text**
**Introduction**

In times of national emergency, bank disappointments, disturbances and arising occurrence of negative genuine premium rates individuals consider gold as perfect resource and like to put investment into such valuable metals. The international investors hunt after a safe investment in the precious metal like gold in the global recession in the history (Opdyke 2010). For instance in current unpredicted worldwide recession related circumstance USA, India and China are front runner in consuming major part of worldwide generated gold. Mpofu (2010) expressed that China, Australia, USA and South Africa separately are huge actors in the worldwide gold creation. Investors invest in gold to hedge against inflation, to counterbalance stock market falloffs and counteract against weakening dollar; whereas, monetary organizations generally use it for security handling for their specialized funded and non-funded products, hence is often considered a confident investment and solid security which can be easily be marketed and sold. At present in Nepal, political government is instable, financial indicators are letting down and stock returns are reducing because of unethical works in stock trade. Along these lines, investors in this extraordinary instability are putting resources into gold fairly than putting resources into stock exchange of Nepal. Therefore demand of gold has been expanded and Nepal has ended up being consumer of gold. Unfortunately, creation of gold in Nepal is nil or low which cannot match the demand of the gold in Nepal. NEPSE which was built up in 1983 is a major stock exchange of Nepal. Yet, in the present circumstance gold has pulled the enthusiasm of investors in light of the fact that there is a bit possibility of showing signs of improvement returns in the stock market investment because of delicate monetary, financial and economy policy practiced in Nepal (Nepal Rastra Bank 2009).

Stock market is considered to be a significant contributor to economic development .The trend related to gold price in Nepal has made investors diversify away from stock market as gold being safe investment while the stock exchange of Nepal has seen a poor or slow progression and considered to be risky investment. The trend is such that while the performance of stock market is poor people are investing in gold in the period of decrease price of gold in international market. The demand of gold is in increasing trend which proves people are highly investing on safe investment like gold, as a result less investment is poured into the stock market in Nepal and also the trend in stock market is unpredictable as it is influenced by several other factors such as inflation, money supply, political environment, etc. This paves to the possibility that demand of gold by investors and stock market performance in Nepal might be related to one another such that the increase in one variable is causing a decrease in another.

In such a case, it could be that the increasing demand of gold is diversifying away investors from the stock market which might have caused the effect in the stock market performance or there possibility that demand of gold is affected by the stock market performance. Thus, it is necessary to understand if there exist a bidirectional causal relationship between Gold and Stock Exchange in Nepal.

Regardless of the assumption and belief of gold impacts stock market performance, only few studies have been made to understand the nature and relationship of Gold price and Stock Exchange performance in Nepal. Some studies and researchers from outside of Nepal points that gold price does have significant impact on stock exchange performance while others point out that there is no substantial impact of gold price on stock market performance. However, research that analyzes the role of gold price on NEPSE performance is rare in Nepal. Even among the few studies made on NEPSE, the impact of gold price on stock market performance has not been studied. Further, the researchers have only taken the macro variables such as Consumer Price Index, inflation, NRB's policy, etc. to capture the relationship with NEPSE performance which does not seize other impact generated on stock market performance through other alternative investments like gold.

Henceforth, this study aims at filling the gaps by assessing the nature and relationship between the Gold price and stock market performance, measured by NEPSE index. The objective of this study is to recognize the causal relationship between Gold and Stock Exchange as well as the elasticity between these variables in the context of Nepal. In specific terms, this study intends to:

A.      Identify the trends of Gold  and Stock Exchange growth in Nepal

B.      Identify relationship and degree of relationship between Gold and NEPSE index.

This Paper examines the impact of gold prices on Nepal Stock Exchange (NEPSE) which is the main stock exchange of Nepal. Gold prices has crossed the limit of  NRs 54269.54 per 10 gram  in August, 2013 in Nepal due to international rising gold prices (Forex 2013).This increasing trend is attracting potential investors in gold investment and diverging current investors from stock investments. In this situation, it is important to examine the concern of this factor gold on NEPSE the major stock market of Nepal. For this purpose my research investigates the relationship between gold prices and Nepal stock exchange performance and its impact on value of Nepse index. For this task the statistical techniques Unit Root Test of Augmented Dickey Fuller (ADF), Phillip Perron, Johansen's Co Integration Test and Granger Causality Test have been used in this paper to demonstrate the relation between average gold price and average NEPSE index. This study is analyzing the long term impact of gold prices on the Nepal stock market.

**Introduction of Nepal Stock Exchange (NEPSE)**

Nepal Stock Exchange is established under the company act, operating under Securities Exchange Act, 1983. The fundamental target of NEPSE is to provide free marketability and liquidity to the administration and corporate securities by encouraging exchanges in its exchanging floor (Nepal Stock Exchange 2015). Introduction of the Company Act in 1964, the first issuance of Government Bond in 1964 and the foundation of Securities Exchange Center Ltd. in 1976 were other huge improvement relating with capital markets.

**Literature Review**

Gulati and Modi (1982) examined that during 1972 to 1982; the gold price were dictated by the interest of gold for investment purposes or by the theoretical interest in gold which is portrayed by monetary exercises because of inflationary expectations and fluctuation of exchange rates. Their study likewise demonstrated that gold costs are exceedingly connected with the investment interest for gold while increment in interest rate lessens the interest for gold venture as it is non interest investment.

Levin and Wright (2006) inspected the relationship between Gold price and the US dollar price. Applying co-integration method on figures from January 1976 to August 2005, long term factors of gold price were set up and a long term relationship between price of gold and the level of U.S. dollar price was found. Study uncovered that the level of U.S dollar price and costs of gold moves together in a measurably in routine fashioned that 1% increase in a U.S dollar price level prompts 1% expansion in gold price; though, if there is an occurrence of any uneven circumstances, this long term relationship is deviated bringing weak relationship. Discoveries of this study additionally investigated there exist a positive relationship between gold price movement and US inflationary rate, US inflation volatility and credit hazard.

Baur and Mc Dermott (2009) led a distinct descriptive and econometric investigation of data from 1979 to 2009. Results of the study demonstrated that gold for the most part is utilized as hedge and considered as safe investment for major European and US stock exchanges however said, these result are not shown in Australia, Canada, Japan and developing markets, for example, BRIS nations. Gold investors utilized it to secure the capital in 'negative economic situations' similar to monetary and financial recession which at present is faced following July 2007 to date in international market like USA, Greece, Spain,etc. This phenomenon made higher interest of gold and a general increment of 75 % is seen in the gold around the world. In addition to this results, Baur and Lucey (2010) explored the relationship of gold price with negative economic situations and discovered curvilinear relationship. They recommended that negative economic situations put important effect on gold investors. Mc Cown and Zimmerman (2006) discovered relationship between nature of gold and expansion supporting; and described gold as Zero beta resource i.e. without economic marked hazard.

Moore and Geoffrey H (1990) inspected relationship between gold price and the estimation value of stock market. The outcomes on empirical idea for the time of 1970 to 1988 investigated a negative relationship between gold price and the value of stock exchanges which showed that an increment in gold price has a tendency to decrease in the value of stock exchanges. These discoveries were additionally acknowledged by who examined the impacts of seven macroeconomic variables (customer value file, currency business sector premium rate, gold cost, modern generation record, oil value, remote swapping scale and cash supply) on the Turkish Capital Markets. Discoveries of his study expressed that Turkish financial investors utilized gold as an optional venture resources; while, in occasion of gold price increment, Turkish investors are diverse away from investment in stocks, resultantly stock prices reduced; therefore, study presented negative relationship between gold price and stock returns (Büyükşalvarcı 2010).

The short-run relation between returns for gold and returns for US stock price indices is little and negative and for a few arrangement and time periods unimportantly not the same as zero. Over the January 1991 and ending in October 2001 period, gold price and US stock price indices are not co integrated. Granger causality tests discover proof of unidirectional causality from US stock returns to returns on the gold price set in the London morning fixing and the closing price. At the cost set toward the afternoon fixing, there is clear proof of pointer between the market for gold and US stocks (Smith 2001).

The relationship between the costs of gold and stock price indices for Armenia over the period starting in January 2005 and ending in December 2010. Over the period inspected, gold price and Armenia stock market are not co-integrated and also gold price and Tehran Stock Exchange (TSE) are not co-integrated despite the fact that the time series of index and gold price didn't have stationary at zero level. Granger causality tests didn't discover confirmation of causality between gold cost and the stock market execution in both nations (Bashiri 2011).

Wang et al (2010) investigated the effects of variances in crude oil price, gold price, and exchange rates of the US dollar price versus different currencies on the stock price indices of the United States, Germany, Japan, Taiwan, and China individually, and in addition the long and short-term correlations among these variables. Observed results demonstrate that there exist co-integrations among variance in oil price, gold price and exchange rate of the dollar versus various currencies, and the securities exchanges in Germany, Japan, Taiwan and China. This shows that there exist long term stable connections among these variables. While there is no co-integration relationship among these variables and the U.S. stock exchange indices which indicates that there is no long term stable relationship among the oil price, gold price and exchange rate and the US stock market indices.

Hasan et al. (2008) explored relationship between KSE and stock markets of developed and developing nations. For this reason, they acquired information of weekly closing value of stock market indices of nine nations for the period of 2000-2006. These nations involved the USA, the UK, France, Germany, Japan, Canada, Italy, Australia and Pakistan. Utilizing Descriptive insights, Correlation lattice, Co mix tests and Granger causality strategy, authors expected to capture the factor of linkages between Karachi stocks market and the stock markets of these other nations for the period 2000 to 2006. Outcomes of their examination gave significant proofs and described that Stock markets of the USA, the UK, Japan and Italy are indicating negative returns in the study period. At the same risk level, negative returns are seen in German stocks markets; while, Karachi stock exchange delineated greatest stocks returns.

Abdalla and Murinde (1997) additionally explored the relationship between the exchange rate and stock prices in India, Pakistan, Korea and Philippines during the period of 1985-1994 by utilizing co-integration statistical method. Muhammad et al. investigated the relationship between exchange rates and stock prices in Pakistan, India, Sri Lanka and Bangladesh for the period 1994-2000 (Muhammad, Rasheed and Fazal 2003). Results of above studies demonstrated that causal relationship among financial variables and Equity Returns don't exist.

In India and Pakistan, Bilal et.al explored and reported that there exist no long-term relationship among Gold prices ,KSE but a significant long-run relationship exist between BSE stock index which are the major stock market in India and Pakistan (Bilal, et al. 2013).

The explored result of Nepal stock market find out NEPSE index has strong positive relationship with inflation and money supply. Benchmark index has negative correlation with interest rate. Domestic stock market quite responsive to macroeconomic condition. NEPSE index influenced by political developments and NRB's policy on margin lending (Shrestha and Subedi 2014).

The presence of causality relationship between stock market and economic development for the year 1988 to 2005 utilizing Granger causality test. The study finds the exact proof of long-term integration and causality of macroeconomic variables and stock market indicators even in a little capital business sector of Nepal. In econometric sense, it defines that the stock market has critical part in deciding economic development and vice versa. The significance of stock market improvement for cultivating economic development in Nepal (G.C 2006). In other study on examined Nepal Stock Exchange demonstrated that the Nepalese stock exchange market is not effective in delicate structure as to the day-of-the-week irregularity however poorly capable as for alternate inconsistencies (K.C. and Joshi 2005).

**Data and Methodology**

**Theoretical Framework**

The recognized conceptual framework assumption for this study is that there is a direct and negative relationship between stock exchange and gold price .This research anticipates and explores two possible ways in which the relationship between these two variables may exist.

The theory assumes that Gold validly represents an alternative investment to stock market by gold price and Stock Exchange of Nepal performance is acknowledged by NEPSE index.

The proposed theoretical framework is illustrated as follows:

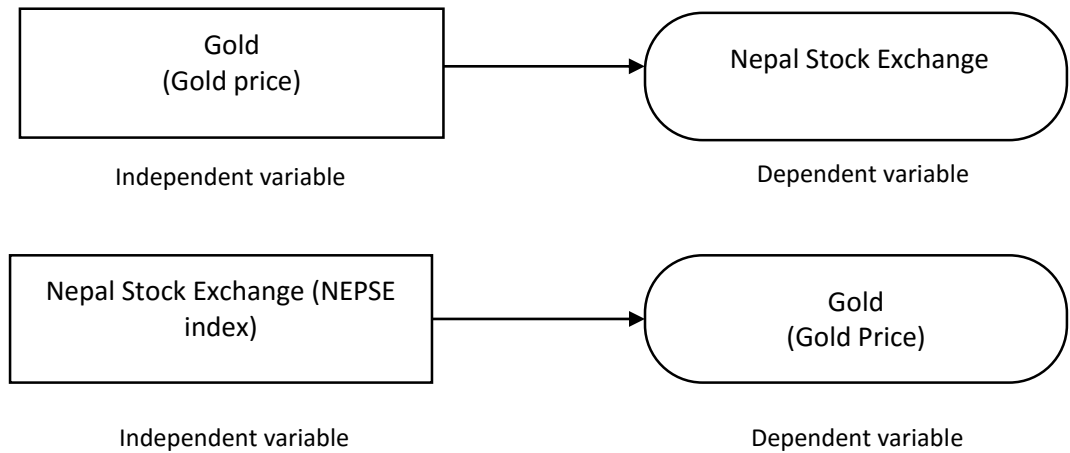

*Figure 1: Theoretical Framework*

**Description of Variables**

The two variables of this study are:

**Gold:** Gold price is taken from the statistics published by Federation of Nepal Gold and Silver Dealers' Association (FENEGOSIDA). The daily basis data is taken which is transformed into average monthly data to confirm the accuracy of used data. The price is in Nepalese Currency i.e. Nepalese Rupees (FENEGOSIDA 2015).

**NEPSE**: NEPSE index is taken from the statistics provided by Nepal Stock Exchange. The daily basis NEPSE index is taken which is transformed into average monthly NEPSE index to ensure the accuracy of NEPSE index (Nepal Stock Exchange 2015).

**Descriptive Statistics and Correlation matrix of Variables**

**Descriptive statistics** are used to evaluate the mean, standard deviation, median, Skewness, etc. of the gold price and NEPSE index.

**Correlation matrix** shows the dependence is any statistical relationship between two random variables or two sets of data. Here correlation matrix is used to show the dependence between gold price and NEPSE index data.

**Statistical Model**

**Background**

The model that has been specified to investigate the causal relationship is:

$GOLD_t = \alpha_0 + NEPSE_t + \acute{\upsilon}_t$ ;……………

$NEPSE_t = \alpha_0 + GOLD_t + \acute{\upsilon}_t$ ;……………………

Where, GOLD: Gold price at time t

NEPSE: NEPSE index at time t

$\acute{\upsilon}$: Error term

t: yearly time period

$\alpha$: constant at time 0

**Hypothesis of the Study**

This paper analyzes the following hypothesis with regard to relationship between Gold and Nepal Stock Exchange Performance…..

$H_1$: Gold price drives Nepal Stock Exchange Performance

$H_2$: Nepal Stock Exchange performance drives Gold price

$H_3$: Bilateral causal relationship which combines $H_1$ and $H_2$.

**Unit Root**

Before developing VAR model, it is required to determine if the both the variables – Gold and NEPSE – are stationary or not. Precisely, stationary means that means and variance of a series are constant throughout the yearly period and the auto-covariance of the series is not time varying. Different methods are available for testing unit root: Dickey Fuller, Augmented Dickey Fuller, Durbin Watson test and Phillip-Perron Test. Here, Augmented Dickey Fuller (ADF) test has been applied for unit root testing. Augmented Dickey Fuller test should result such that the variables are non-stationary in their level, however the variables should be stationary after first differencing.

**VAR model**

In order to test the mentioned hypothesis, two variable VAR model is developed using as follows (Oh, 2005):

$$\begin{bmatrix} \Delta GOLD(t) \\ \Delta NEPSE(t) \end{bmatrix} = \alpha 0 + \alpha 1 \begin{bmatrix} \Delta GOLD(t-1) \\ \Delta NEPSE(t-1) \end{bmatrix} + \alpha 2 \begin{bmatrix} \Delta GOLD(t-2) \\ \Delta NEPSE(t-2) \end{bmatrix} + \cdots \alpha p \begin{bmatrix} \Delta GOLD(t-p) \\ \Delta NEPSE(t-p) \end{bmatrix} + Ut\ldots\ldots\ldots\ldots\ldots\ldots\ldots\ldots\ldots$$

Where, $\alpha_0$: vector of constant term

$\alpha_i$: matrix of parameters

$U_t$: innovation term

**Lag selection**

The optimum no. of lag is selected using Schwartz Bayesian Criteria (SBC). The equation for SBC is:

$$SBC = T \, log|\textstyle\sum x| + N \log(T) \ldots\ldots\ldots$$

Where, $|\sum x|$: determinant of variance/covariance matrix of the residuals from (3)

N: Total number of parameters estimated in all equations

T: Number of time points

**Johansen's Co-integration Test**

If the variables contain unit root then co-integration between variables are examined using Johansen's Co-integration Test (Johansen 1991) : According to Granger, "co-integration between variables means that non-stationary variables are integrated in the same order with residuals stationary. If there is co-integration between two variables, there is a long-run effect that prevents two time series from drifting away from each other and there exists a force to converge into long-run equilibrium" (C. Granger 1969).

Two test statistics, Trace statistics and Maximum Eigenvalue Test statistics are used to identify the number of co-integrating vectors. The trace test statistic for the null hypothesis that there are at most r distinct co-integrating vectors is (Johansen 1991) :

$$\lambda_{trace} = T \sum_{J=r+1}^{n} (1 - \lambda_J)$$

Where, $\lambda_J$: n – r smallest squared correlation between $X_{t-k}$ and $\Delta X_t$ where Xt = (NEPSE$_t$ or GOLD$_t$).

The maximum likelihood ratio statistic for testing null hypothesis of at most r co-integrating vectors against the alternative hypothesis of r+1 co-integrating vectors, maximum eigenvalue statistic, is (Ekanayake, 1999):

$$\lambda_{max} = -T \ln(1 - \lambda_{r+1})$$

**Granger Causality Test**

The three hypothesis as mentioned above – gold price impacts NEPSE performance, NEPSE performance impacts gold price and bilateral causal relationship – are tested using Granger Causality Approach. This method is best implemented when lags of one variable enters into equation for another variable (Enders, 2004).

The equation system from (3) is considered in following way to test for Granger Causality:

$$\Delta GOLD_t = \alpha_1 + \sum_{p=1}^{n} \beta_{1p} \, \Delta NEPSE_{(t-p)} + \sum_{p=1}^{n} \delta_{1p} \, \Delta GOLD_{(t-1)} + \gamma_1 (D_{t-n-1}) + \varepsilon_{1t}$$

$$\Delta NEPSE_t = \alpha_2 + \sum_{p=1}^{n} \beta_{2p} \, \Delta GOLD_{(t-p)} + \sum_{p=1}^{n} \delta_{2p} \, \Delta NEPSE_{(t-1)} + \gamma_2 (D_{t-n-1}) + \varepsilon_{2t}$$

Where, $\alpha 1, \alpha 2$: Vector of constant terms

GOLD$_t$ : Gold price at time t

NEPSE$_t$ : NEPSE index in time t

t: Time period based on year

p: optimum lag

$\beta 1p$: measure of influence of NEPSE$_{(t-p)}$ on GOLD$_t$

$\beta 2p$: measure of influence of GOLD$_{(t-p)}$ on NEPSE$_t$

$\varepsilon 1t$, $\varepsilon 2t$: random error

From, the above equations, three hypothesis of the study are tested as follows:

Hypothesis (I)

Null Hypothesis: NEPSE index in Nepal doesn't granger cause Gold price of Nepal

Alternate Hypothesis: NEPSE index in Nepal granger causes Gold price of Nepal

Hypothesis (II)

Null Hypothesis: Gold price of Nepal doesn't granger cause NEPSE index in Nepal

Alternate Hypothesis: Gold price of Nepal granger causes NEPSE index in Nepal.

Hypothesis (III)

Null Hypothesis: There is no two way relationship between Gold price and NEPSE index arrivals in Nepal

Alternate Hypothesis: There is two way relationship between Gold price and NEPSE index in Nepal

**Collection Tools**

The data collected is completely based on secondary source. The information and data regarding gold price is collected through FENEGOSIDA on their official website (FENEGOSIDA 2015). Likewise, the information and data regarding NEPSE of Nepal is collected from the Nepal Stock Exchange Statistics provided by NEPSE on their official website (Nepal Stock Exchange 2015). This study is comprised with the period from 2010:09 to 2015:06 of gold price and NEPSE index data. Data analysis is done using statistical software like MS-Excel, EVIEWS and R.

**Empirical Results:**

**Descriptive Statistics:**

Descriptive statistics for Nepal Stock Exchange and average Gold prices are given in the Table 1. This includes the description of mean, standard deviation, kurtosis, skewness and variance etc. A careful study shows that average Gold prices have Mean value of NRs 44326.77 and Nepal Stock Exchange has a Mean value of 575.6215 and average Gold prices has a standard deviation of 5095.400 and Nepal Stock Exchange of 237.9882.

*Table 1: Summary Statistics*

| Statistics | GOLD | NEPSE |
|---|---|---|
| Mean | 44326.77 | 575.6215 |
| Median | 45349.41 | 500.2166 |
| Maximum | 51683.33 | 1034.810 |
| Minimum | 32610.00 | 308.4009 |
| Std. Dev. | 5095.488 | 237.9882 |
| Skewness | -1.010310 | 0.598399 |
| Kurtosis | 3.137694 | 1.789171 |
| Jarque-Bera | 9.912850 | 7.004545 |
| Probability | 0.007038 | 0.030129 |
| Sum | 2570952. | 33386.05 |
| Sum Sq. Dev. | 1.48E+09 | 3228388. |
| Observations | 58 | 58 |

**Correlation Matrix:**

Table 2 shows the results of correlation matrix and it indicates that Nepal Stock Exchange (NEPSE Index) is positively correlated with average Gold prices because this positive relation is not much strong as the value lies as 0.190415 for correlation with Average Gold Prices and Nepal Stock Exchange (NEPSE).

*Table 2: Correlation between Gold price and NEPSE index*

| | GOLD | NEPSE |
|---|---|---|
| GOLD | 1.000000 | 0.190415 |
| NEPSE | 0.190415 | 1.000000 |

Correlation analysis is not a strong technique because it does not discuss the cause and effect relationship. In order to know a clear picture of the relationship we perform Co Integration analysis that tests the co-movement of NEPSE index and average Gold prices.

**Unit Root Analysis:**

**Unit root analysis (Augumented Dickey Fuller Test)**

In order to run the Co Integration analysis the series should be stationary. So, in the first step, the stationary of the series has been tested. Table 3 reveals that time series are not stationary at levels. However, Table shows that the NEPSE and average Gold price series are stationary at 1st Difference. Augmented Dickey Fuller Unit Root Analysis test reveals that errors have constant variance and are statistically independent.

*Table 3: Unit Root Analysis (ADF test) of gold price and NEPSE index*

| | Augmented Dickey Fuller (Level) | p-value | Augmented Dickey Fuller (1st Difference) | p-value |
|---|---|---|---|---|
| Gold Prices | -2.3977 | 0.1469 | -7.245509 | 0 |
| Nepal Stock Exchange | 0.138162 | 0.9660 | -5.649221 | 0 |
| Critical Values | | | | |
| 1% | -3.550396 | | -3.552666 | |
| 5% | -2.913549 | | -2.914517 | |
| 10% | -2.594521 | | -2.595033 | |

**Unit Root Analysis (Phillip-Perron):**

An alternative test of Unit Root Tests i.e. Phillip Perron test is used to check the stationary of the data. This test allows the error variance to be heterogeneously distributed and less dependent. Table 4 proves the results of previous test (table-3) that all series are stationary at 1st Difference. So, Co Integration test can be applied on these variables.

*Table 4: Unit Root Analysis (Phillip-Perron test) of gold price and NEPSE index*

| | Phillips-Perron (Level) | p-value | Phillips-Perron (1st Difference) | p-value |
|---|---|---|---|---|
| Gold prices | -2.397375 | 0.147 | -7.245509 | 0 |
| NEPSE | -0.019373 | 0.9525 | -5.489734 | 0 |
| Critical Values | | | | |

| 1% | -3.550396 | | -3.552666 | |
| 5% | -2.913549 | | -2.914517 | |
| 10% | -2.594521 | | -2.595033 | |

**Vector Auto Regression**

The table 5 shows selection order criteria SBIC (Schwarz Bayesian information criterion) and AIC (Akaike information criterion) both to choose the lag for cointegration test. The result shows that lag 1 must be chosen to continue further for Johansen Co-integration test.

*Table 5: VAR Lag Order Selection Criteria*

| Estimate VAR at Lag 0 | |
|---|---|
| Akaike information criterion | 33.28099 |
| Schwarz criterion | 33.35534 |
| Estimate VAR at Lag 1 | |
| Akaike information criterion | 27.84157* |
| Schwarz criterion | 28.06462* |
| Estimate VAR at Lag 2 | |
| Akaike information criterion | 27.92118 |
| Schwarz criterion | 28.29293 |
| Estimate VAR at Lag 3 | |
| Akaike information criterion | 28.02428 |
| Schwarz criterion | 28.54474 |
| Estimate VAR at Lag 4 | |
| Akaike information criterion | 28.16128 |
| Schwarz criterion | 28.83044 |
| Estimate VAR at Lag 5 | |
| Akaike information criterion | 28.17852 |
| Schwarz criterion | 28.99638 |

**Johansen Co Integration test:**

Johansen Co Integration Test (Trace Statistics and Max Eigen) is used for analyzing the co integrating vectors between Nepal Stock Exchange and average Gold price time series. It shows a long term relationship among dependent variable (Average NEPSE) and independent variable (Average Gold price).

**Bivariate Co Integration Analysis Trace Statistics**

Results of this analysis (Trace Statistics) are given in the **Table 6**. This assures long term relationship does not exist among the variables at the 0.05 level.

*Table 6: Bivariate Co Integration Analysis Trace Statistics*

| Hypothesized No. of CE(s) | Eigenvalue | Trace Statistic | 0.05 Critical Value | Prob.** | Remarks |
|---|---|---|---|---|---|
| None | 0.169895 | 11.26272 | 15.49471 | 0.1958 | No Co Integration |
| At most 1 | 0.014806 | 0.835353 | 3.841466 | 0.3607 | |

**Bivariate Co Integration Analysis Max-Eigen Value Statistics:**

Max-Eigen value statistics test is used to confirm the results of co integration Trace statistics analysis. It also ensures two co integrating equations have no level of significance at 0.05 level showing no co integration among two variables. Results are given in the table 7.

*Table 7: Bivariate Co Integration Analysis Max-Eigen Value Statistics*

| **Hypothesized No of CE(s)** | **Eigenvalue** | **Max-Eigen Statistic** | **0.05 Critical Value** | **Prob.** | **Remarks** |
|---|---|---|---|---|---|
| None | 0.169895 | 10.42736 | 14.26460 | 0.1854 | No Co integration |
| At most 1 | 0.014806 | 0.835353 | 3.841466 | 0.3607 | |

**Granger Causality Analysis:**

If Co Integration exists between the variables then there must exist a paired Granger-Causality between the variables. Table 6 shows that long term relationship does not exist among the variables. So, Granger- Causality test cannot apply on these variables because it cannot determine the direction of causality in the right way. However the known result of Granger-Causality is tested and presented below in Table 8.

*Table 8: Granger Causality Analysis between gold price and NEPSE index*

| Null Hypothesis: | Obs | F-Statistic | Prob. |
|---|---|---|---|
| NEPSE does not Granger Cause GOLD | 57 | 1.38296 | 0.2448 |
| GOLD does not Granger Cause NEPSE |  | 2.35891 | 0.1304 |

## Summary, Discussions, and Implications

In conclusion, we can say that there exist no relationship between monthly average gold prices and monthly average NEPSE index. These results depict no long term relationship. In this study we applied Unit Root Augmented Dickey Fuller Test for the Stationary time series trend. When we applied Correlation test, it proved that the average gold price had a positive relationship with average NEPSE index which was not strong relationship. In another test, we developed VAR model which applied Co-Integration Test on the basis of monthly data to analyze the long term relationship between gold prices and NEPSE index. But from this test it was proved that there was no existence of long term relationship between these two variables. In this model, we used Lag length 1 given by Schwarz information criterion for analyzing the effects along the period of time. Final test, Granger causality test cannot apply on these variables because Co Integration does not exist between the gold prices and NEPSE index from Johansen Co Integration test. However, Granger Causality test also showed and made clear there exist no causal relationship among gold price and NEPSE index. In conclusion based on our research, we can say that gold price does not have impact on stock exchange and vice versa.

Derivatives such as gold forwards, futures and options have become very popular and have been traded on various exchanges around the world and over-the-counter directly to the private market. In the USA, gold futures are primarily traded on the New York Commodities Exchange. Perhaps, this explains the co-movement of gold prices and stock prices in the outcome of global financial crisis. Findings of this study provide significant insights for academician, researchers and for local and international investors especially who are interested to invest in capital market for prudent decision making.

Other researchers can study these two variables (i.e. gold prices and NEPSE) in short term relationship and impact of other macro variables on NEPSE. More studies can also be conducted by using the average Gold prices in ounce or tola rather in grams and effect of gold prices on Stock Exchanges of other South Asian country can be examined. So other factor responsible for the stock exchange performance can be viewed and understood more clearly. Such research and findings will help decision makers to make appropriate decision based on the information available.